\documentclass[preprint]{aastex}

\begin{document}

%\begin{flushleft}
%{\small\bf v21; Sept 6, 2008}
%\end{flushleft}

\title{Pulsar timing and spacetime curvature}
\author{Teviet Creighton}
\author{Fredrick A.~Jenet}
\author{Richard H. Price}
\affil{Center for Gravitational Wave Astronomy and
Department of Physics and Astronomy, University of Texas at Brownsville,
Brownsville, Texas 78520
}

\begin{abstract}
We analyze the effect of weak field gravitational waves on the timing
of pulsars, with particular attention to gauge invariance, that is, to
the effects that are independent of the choice of coordinates. We
find: (i)~the Doppler shift cannot be separated into gauge invariant
gravitational wave and kinetic contributions; (ii)~a gauge invariant
separation {\it can} be made for the time derivative of the Doppler
shift in which the gravitational wave contribution is directly related
to the Riemann tensor, and the kinetic contribution is that for
special relativity; (iii) the gauge dependent effects in the Doppler
shift play no role in the program of gravitational wave detection via
pulsar timing. The direct connection shown between pulsar timing and
the Riemann tensor of the gravitational waves will be of importance in
discussions of gravitational waves from alternative
(non-Einsteinian) theories of gravitation.
\end{abstract}

\keywords{pulsars:general, gravitational waves}

\section{Introduction}\label{sec:intro} 

The possibility of detecting gravitational waves through pulsar
timing, first suggested independently by \cite{sazhin78} and \cite{Detweiler79}, is of increasing interest as part of the
birth of gravitational wave astronomy, and the details of this
technique continue to be advanced by several researchers \citep{HD83,backerhellingsAnnRev86,hellingsAJ86,ktr94,LB01,JHLM05,Hobbs08}.
The basis of this detection technique is the effect that a
gravitational wave has on the arrival times of pulsar signals.  In the
analysis of this effect \citep{hellingsPRD81,backerhellingsAnnRev86},
the gravitational wave has been described in terms of perturbations
$h_{\mu\nu} \equiv g_{\mu\nu}-\eta_{\mu\nu} $ of the spacetime metric
$g_{\mu\nu}$. This description of gravitational waves is analogous to
describing electromagnetic waves using the potentials $\{\Phi,{\bf A}
\}$. In particular, coordinate transformations are analogous to the
gauge transformations of electromagnetic theory, and -- at least in
linearized general relativity -- are also called gauge
transformations.

If electromagnetic fields are to be described in terms of potentials,
then one must choose a particular gauge for the description. Similarly
in relativistic gravity, coordinate conditions (``gauge fixing'') must
be added.  The need for a gauge choice gives rise to the possibility
of confusing a gauge effect with a physical effect.  This confusion,
for relativistic gravity, kept it a controversial question for many
years whether gravitational waves were physical, or were coordinate
waves \citep{KEN07}. A further source of confusion has been the
distinction between the Doppler shift due to gravitational waves, and
the Doppler shift due to relative motion of the emitting pulsar and
the receiving telescope.  This has the potential to be particularly
confusing, since the relative motion is a manifestation of spacetime
curvature, as is the gravitational wave.

Our purpose here is to clarify the description of pulsar timing, by
analyzing pulsar timing with particular attention to what aspects of
the pulsar Doppler shift are gauge dependent and what aspects are
physical. What we will show is that pulsar timing can be analyzed in a
way in which there is an incontrovertible contribution due to
gravitational waves, and a contribution due to relative motion of
emitter and receiver. The result is
couched completely in terms of quantities that are
intuitively appealing, as well as mathematically well defined. The
gravitational wave contribution will be based on the Riemann tensor
which is gauge invariant very much as are the electric and magnetic
fields in classical electromagnetism.  We will see, moreover, that in
the literature there are claims about Doppler shift effects that are
not gauge invariant, but that these statements refer to intermediate
steps in the analysis of pulsar timing and do not affect the overall
program of pulsar timing. The technique of gravitational wave
detection by pulsar timing is physically valid; it is cleanly
distinguishable both from coordinate effects and from source/receiver
motions.

Besides clarifying what is and is not physically meaningful in pulsar
timing, this paper helps to establish a firm background for studying the
nonstandard polarization modes of alternative theories of gravitation,
modes for which pulsar-timing detection may be particularly useful. 
Such modes are most easily described in terms of components of the Riemann
tensor, not in terms of metric perturbations.

The paper is organized with Sec.~\ref{sec:gaugeinv} as its
mathematical heart.  In that section we derive the expressions that
form the basis of our analysis, and that show clearly what is and what
is not gauge invariant in pulsar timing.  To help focus on the
distinction between kinematic and gravitational effects, in
Sec.~\ref{sec:gaugeinv} we assume that the motions of the emitter and
the receiver of pulses are driven by nongravitational forces. We
remedy this unphysical assumption in Sec.~\ref{sec:pngravity}, where
we add gravity as the source of the astrophysical motions of the
emitter and receiver.  In Sec.~\ref{sec:conc} we discuss the
implications of these results for the way in which pulsar-timing
gravitational wave detection is to be carried out, and is to be
viewed. In the Appendix we fill in some details of
Sec.~\ref{sec:gaugeinv} that we thought would divert too much
attention from the main point of that section.

We have chosen to present these results with a minimum of unnecessary
generality that would make the mathematics more elegant, but more
obscure.  Except as noted, we use the notation and conventions of the
text by \cite{MTW}. 

\section{Doppler shift and gauge invariance}\label{sec:gaugeinv} 

The Doppler shift from emitter to observer can be considered to have
two sources: (i)~general relativisitic: the effects of gravitational
waves, and (ii)~ special relativistic: the relative motion of emitter
and receiver. The second effect will be much larger than the first in
astrophysical situations.  The relative velociy of astrophysical
bodies is on the order of $v\sim10^{-3}$ (in $c=1$ units), while the
characteristic magnitude of gravitational waves is smaller than
$10^{-20}$, usually much smaller.  In our development we keep special
relatistic effects to all orders of $v$, but we will ignore effects of
order $v|h_{\mu\nu}|$. The justification is that gravitational effects
will be at the limit of detectibility; effects that are smaller by
$10^{-3}$ are not of immediate interest. Here and below we will use
$v$ to denote an astrophysical velocity, and we will repeatly use the
fact that $v|h_{\mu\nu}|$ can be ignored.

As pointed out in Sec.~\ref{sec:intro}, here we will make the
artificial assumption that the emitter and receiver are point
particles that are being driven in accelerated motions by
nongravitational forces, such as rocket engines.  In the current
section, then, the only curvature of spacetime is due to gravitational
waves.

We will denote our emitter worldline as E, and receiver worldline as
R.  We choose Minkowski background coordinates $t,x,y,z$ for the
background so that they are appropriate for the approximations just
discussed. That is, in these coordinates the E and R worldlines are at
rest aside from velocities of order $v$, and the metric perturbations
$h_{\mu\nu}$ due to gravitational waves are extremely small.
The 4-momentum of a photon from emission to reception is written
as $P^\mu=P^0(1,\vec{n})$, so that $\vec{n}$ plays the role of a unit
vector pointing (in the Minkowski background) from the emitter to the receiver.

We now consider the following expression
\begin{equation}\label{DoppwithU}
\mbox{Dopp}=- \int_{\rm E}^{\rm R}
\left({\textstyle\frac{1}{2}}\,h_{tt,t}
+n^jh_{tj,t}+{\textstyle\frac{1}{2}}n^jn^k\,h_{jk,t}\right)\,d\lambda
+\left[U^{t}-n_kU^{k}\right]^R
-\left[U^{t}-n_kU^{k}\right]^E
\,.
\end{equation}
Here $U^\mu$ is the 4-velocity of of the emitter (E) and receiver (R),
at the events of emission and reception; $d\lambda$ indicates
integration along the photon worldline, with $dt=d\lambda$ and
$dx^j=n^jd\lambda$; Latin indices are spatial (referring to the
$x,y,z$ components of the coordinate basis).

In the Appendix we show that the expression in Eq.~(\ref{DoppwithU})
represents the Doppler shift of the photon, that is, the fractional 
difference by which 
the photon energy observed by the receiver is greater than that
observed by the emitter.  Here we focus on the gauge property of the
expression, the changes induced in the expression by a coordinate
transformation $x^{\mu\ {\rm new}}= x^{\mu}+\xi^\mu $ in which the
gauge vector $\vec{\xi}$ is of the order of the metric perturbations
$h_{\mu\nu}$.

The standard gauge transformations of the perturbations and of the
components of the 4-velocities are
\begin{equation}\label{gaugexform} 
h_{\mu\nu}^{\rm new}=
h_{\mu\nu}+\xi_{\mu,\nu}+\xi_{\nu,\mu}
\quad\quad\quad
U^{\mu\ \rm new}=
\xi^\alpha
U^\mu
_{\ ,\alpha}-U^{\alpha}\xi^\mu_{\ ,\alpha}
\,.
\end{equation}
We now note that the 4-velocities have the components
$U^\mu=\delta^\mu_t$ aside from corrections of order $v$. With terms of
order $v|\vec{\xi}|$  ignored we are left with
\begin{equation}\label{Uxform2} 
U^{\mu\ \rm new}=U^{\mu}-\xi^{\mu}_{,t}\,.
\end{equation}
We also note that in Eq.~(\ref{DoppwithU}) a gauge change in $\vec{n}$
would give terms of order $v|\vec{\xi}|$ and
$h_{\mu\nu}|\vec{\xi}|$, so no gauge transformation of $\vec{n}$
is carried out.

A straightforward calculation shows the gauge invariance of Dopp: 
\begin{displaymath}
\delta\mbox{Dopp}\equiv
\mbox{Dopp}^{\rm new}-\mbox{Dopp}
=
- \int_{\rm E}^{\rm R}
\left(\xi_{t,t,t}+n^j\xi_{t,j,t}+n^j\xi_{j,t,t}+n^jn^k\xi_{j,k,t}
\right)\,d\lambda
+\left[\xi_{t,t}+n^j\xi_{j,t}\right]^{\rm R}
-\left[\xi_{t,t}+n^j\xi_{j,t}\right]^{\rm R}
\end{displaymath}\begin{equation}
=
- \int_{\rm E}^{\rm R}
\left(\frac{\partial}{\partial t}+n^p\frac{\partial}{\partial x^p}
\right)\left(\xi_{t,t}+n^q\xi_{q,t}\right)
\,d\lambda
+\left[\xi_{t,t}+n^j\xi_{j,t}\right]^{\rm R}
-\left[\xi_{t,t}+n^j\xi_{j,t}\right]^{\rm R}
=0\,.
\end{equation}
(It is worth noting here that time derivatives of $\vec{n}$ can be
ignored since these time derivatives are of order $v$ and would be
multiplied by terms of order $|\vec{\xi}|$.)

Since the Doppler shift, as defined, refers to an objective physical
measurement, the fact that it is gauge invariant is simply a
consistency requirement. The details of the gauge transformation,
however, underscore an important point: neither the integral nor the
4-velocity contributions to Eq.~(\ref{DoppwithU}) is separately gauge
invariant. Thus, the temptation to identify the integral in
Eq.~(\ref{DoppwithU}) as the gravitational wave contribution and the
4-velocity terms as the kinetic contribution must be avoided,
since that identification has no invariant meaning.

To arrive at a more physically useful expression we take the derivate
of Eq.~(\ref{DoppwithU}) with respect to the coordinate time $t$.
(In doing this we note again that $d\vec{n}/dt$ is of order $v$, and hence 
that time differentiation of $\vec{n}$ in the integral can be ignored.)
The result of time differentiation is:
\begin{eqnarray}\label{dDoppdt}
\frac{d\,\mbox{Dopp}}{dt}&=&-\int_{\rm E}^{\rm R}
\left({\textstyle\frac{1}{2}}\,h_{tt,tt} +
n^jh_{tj,tt}+{\textstyle\frac{1}{2}}\,n^jn^kh_{jk,tt}\right)\,d\lambda
\\ \nonumber
&+&\left[\frac{dU^{t}}{dt}-n_j\frac{dU^{j}}{dt}-\frac{dn_j}{dt}U^j\right]^R
-\left[\frac{dU^{t}}{dt}-n_j\frac{dU^{j}}{dt}-\frac{dn_j}{dt}U^j\right]^E\,.
\end{eqnarray}
A few comments here on the time differentiation are appropriate.  The
total time derivative of the integral should, in principle, include
the change in the integral due to the change of the  time of the end points
of the integral. But this change involves $v$, and hence terms of
order $v|h_{\mu\nu}|$ which we ignore.  The time changing endpoints,
on the other hand, cannot be ignored in the 4-velocity terms, since
these terms are not multipled by metric perturbations.  The time
derivatives $d/dt$ in the 4-velocity terms are therefore understood to
be the derivatives along the worldlines, i.e.,
${d}/{dt}={\partial_t}+v^k{\partial_{x^k}}$.

Next we consider the  components of the 
4-acceleration ${a^\mu}$ of the E and R worldlines
\begin{eqnarray}
a^j=\gamma\,\frac{dU^j}{dt}
+\left(U^{t}
\right)^2\Gamma^j_{tt}&=&\gamma\,\frac{dU^j}{dt} +h_{jt,t}
-\textstyle{\frac{1}{2}}h_{tt,j}\\
a^t=\gamma\,\frac{dU^t}{dt}
+\left(U^{t}
\right)^2\Gamma^t_{tt}&=&\gamma\,\frac{dU^t}{dt}
-\textstyle{\frac{1}{2}}h_{tt,t}
\end{eqnarray}
where we have ignored terms of order 
$v|h_{\mu\nu}|$ in the $\Gamma
$ term, and where 
$\gamma
$, as usual, represents $1/\sqrt{1-v^2
\;}\,.
$
When these equations are used in  Eq.~(\ref{dDoppdt}) we get
\begin{eqnarray}
\frac{d\,\mbox{Dopp}}{dt}&=&\int_{\rm E}^{\rm R}
\left({-\,\textstyle\frac{1}{2}}\,n^jn^kh_{tt,jk}+
n^jn^kh_{tj,tk}-\,{\textstyle\frac{1}{2}}\,n^jn^kh_{jk,tt}\right)\,d\lambda
\\ \nonumber
&+&\left[\frac{a^t-n_ja^j}{\gamma}\right]^R
-\left[\frac{a^t-n_ja^j}{\gamma}\right]^E
+\left[-U_j\,\frac{dn^j}{dt}\right]^R
-\left[-U_j\,\frac{dn^j}{dt}\right]^E
\end{eqnarray}
\begin{equation}\label{dDoppdt2}
%\frac{d\,\mbox{Dopp}}{dt}
=\int_{\rm E}^{\rm R}
n^jn^kR_{tjtk}d\lambda
+\left[\frac{a^t-n_ja^j}{\gamma}\right]^R
-\left[\frac{a^t-n_ja^j}{\gamma}\right]^E
+\left[-U_j\,\frac{dn^j}{dt}\right]^R
-\left[-U_j\,\frac{dn^j}{dt}\right]^E\,,
\end{equation}
where $R_{tjtk}$ represents the components of the Riemann tensor.

The expression in Eq.~(\ref{dDoppdt2}) is, of course, gauge invariant,
but unlike the gauge invariant expression in Eq.~(\ref{DoppwithU}),
the individual contributions are now gauge invariant. The integrand
contains only a projection of the manifestly gauge invariant Riemann
tensor. The 4-velocity and 4-acceleration terms are all of first or
higher order in $v$, so their gauge changes would be of order
$v|h_{\mu\nu}|$, and hence ignorable.  Unlike the the expression in
Eq.~(\ref{DoppwithU}), for the Doppler shift, the expression in
Eq.~(\ref{dDoppdt2}) for the {\it time derivative} of the Doppler
shift contains contributions that have physical 
gauge-invariant meaning; the integral gives the effect of
gravitational waves and the remaining terms give the special
relativistic contributions due to acceleration (the 4-acceleration
terms) and to the relative geometry of the worldlines (the $d\vec{n}/dt$
terms).

\section{Gravitationally driven orbits}\label{sec:pngravity} 

In Sec.~\ref{sec:gaugeinv} we assumed that the motions of the emitter
and receiver were driven by nongravitational forces (``rocket
engines''), so that the role of spacetime curvature lay solely in the
perturbations $h_{\mu\nu}$ identified as gravitational waves. In
reality, of course, astrophysical orbital motions are driven by
gravity. In this section we explain how to incorporate other aspects
of gravity, in particular, orbital forces, into the considerations of
Sec.~\ref{sec:gaugeinv}.

We imagine now that the emitter and receiver are on astrophysical
bodies that are moving under the influence of other astrophysical
bodies.  We have already assumed in Sec.~\ref{sec:gaugeinv} that
orbital velocities are small compared to $c$.  This means that the
gravitational interactions among all bodies are weak-field
interactions. More precisely, for our system of gravitationally
interacting bodies, the Newtonian potential $\Phi$ due to one body, at
the position of the other body, must be small compared to $c^2=1$.  We
now make the additional assumption that the gravitational field is
weak ($|\Phi|\ll1$) everywhere along the photon path.  This allows us
to treat all non-gravitational-wave fields relevant to the problem as
being adequately described by a post-Newtonian (pN)
approximation \citep{will}. Though we may consider a higher order pN
approximation, we do not consider orders high enough (order $(v^7)$)
for the orbiting bodies themselves to become sources of gravitational
radiation.

In a pN approximation, a background Minkowski-like coordinate system
is used, and metric perturbations in terms of these potentials are
required to have ``post-Newtonian'' character \citep{will}.  This
character requires, among other things, that 
the potentials be
functions only of the coordinate separation of source point and field
point. The equations determining the metric perturbations are computed
from Einstein's equations truncated to the order of the pN
approximation being used. Gravitationally driven motions are then
taken to be the geodesics of the pN metric, and may be interpreted as
having accelerations in the flat background.

Due to the nature of the pN approach, a pN gauge vector $\xi_\mu$ would
have to have pN character. This very strong constraint turns out to
leave almost no gauge freedom.  Moreover, there is a standard pN
coordinate system~\citep{will} in which even the small gauge freedom is
removed.  When this gauge is adopted, the question of pN
gauge transformations ceases to exist. 

Gauge transformations can still be made, of course. The $h_{\mu\nu}$
gravitational wave metric perturbations do {\em not} have pN
character.  So long as our gauge field $\xi_\mu$ is not pN in
character it is clearly separable from any choice of pN gauge, and
does not affect the construction of pN coordinates.  Again we note
that this presumes that we ignore any contribution to the $h_{\mu\nu}$
field of gravitational waves from the pN-modelled bodies themselves.

With these considerations we can conclude that the gauge
transformations of Sec.~\ref{sec:gaugeinv} can be repeated in the pN
plus gravitational wave spacetime, with the accelerations taken as
those (relative to the Minkowski background) coming from pN
theory. One might worry that the gravitational wave gauge
transformations in Sec.~\ref{sec:gaugeinv} were made relative to a
flat background, while the pN spacetime is not flat. But this simply
means that we are ignoring terms of order $|\Phi||\xi|$, which are
analogous to the terms of order $v^2|\xi|$ that we ignored in
Sec.~\ref{sec:gaugeinv}.

Though we are primarily interested here in gauge questions, the matter
of the pN metric perturbations raises an important separate issue. The
photon's Doppler shift will be affected by the surface gravitational
field of the emitter and receiver. In addition, if the the photon
happens to pass close to a another astrophysical body the pN fields of
that other body will affect pulsar time-of-arrivals by altering photon
path lengths and through the Shapiro time delay \citep{shapiroTD}.
These effects can be separately evaluated (as part of a pN
calculation) and added to the Doppler shift calculations of
Sec.~\ref{sec:gaugeinv}. Since the effects are small, they do not
``mix'' with the $h_{\mu\nu}$ terms.

We close this short section by pointing out that there are limits to
the clean separation of kinematic and gravitational wave terms.  If
the gravitational fields are strong, if orbital velocities are
relativistic, or gravitational potentials along the photon path are of
order unity, then the gauge-invariance demonstration in
Sec.~\ref{sec:gaugeinv} fails to hold. Indeed, it is intuitively
appealing that in such a case it should not be possible to make a
general distinction between small perturbations of spacetime and
motions of astrophysical bodies. It would sill be possible, of course,
to make a practical distinction between the two if the gravitational
waves had a significantly higher frequency than the time scale of the
kinematics.

\section{Conclusions}\label{sec:conc} 

A physical measurement, such as the Doppler shift of a photon, is not
changed by the changes in our mathematical choices, so the gauge
invariance of the expression in Eq.~(\ref{DoppwithU}) is simply a
check of consistency. What {\it is} subject to our mathematical
choices is the interpretation of the terms in Eq.~(\ref{DoppwithU}),
in particular the interpretation of the integral term as the
gravitational wave contribution. This interpretation is clearly
incorrect since we could, for example, choose coordinates that makes
the integral vanish for any photon, and put all of the Doppler shift
into the kinetic term.

Though integrals like that in Eq.~(\ref{DoppwithU}) have appeared in
the literature \citep{Detweiler79,MAS82,BCR83} in discussions of pulsar
timing, in practice, this causes no real difficulty.  The Doppler
shift can be understood to be the time integral of the gauge invariant
expression in Eq.~(\ref{dDoppdt}).  That time integral will contain an
integration constant that cannot be meaningfully separated into
gravity and kinematics. The observed phase of the arriving pulses, the
raw data of pulsar timing, will be the next time integral, and will
contain a term $A+Bt $, where $A$ and $B$ are such integration
constants. These integration-constant terms play no role in the actual
analysis of timing residuals. The program of gravity wave detection by
pulsar timing, therefore, has a solid physical foundation based on the
Riemann tensor.

This relationship of timing residuals to the Riemann tensor has a
useful secondary benefit. As pointed out in the introduction, pulsar
timing has the potential to be a particularly sensitive probe of
non-Einsteinian polarization modes of gravitational
waves \citep{leejenetprice}. The description of these modes 
uses the components of the Riemann tensor. The expression in
Eq.~(\ref{dDoppdt}) gives a clear and unequivocal description of how
these nonstandard gravitational waves affect pulsar timing residuals.

\section{Acknowledgment} We gratefully acknowledge support by the
National Science Foundation under grants AST0545837 and PHY0554367. We
also thank the NASA Center for Gravitational Wave Astronomy at
University of Texas at Brownsville. We thank Kejia Lee for useful 
discussions of this work.

\section*{Appendix: Derivation of the gauge invariant expression for 
the Doppler shift} 

%%%%%%%%%%%%%%%%%%%%%%%%%%%%%%%%%%%%%%%%%%%%%%
\begin{figure}[ht]
\begin{center}
\includegraphics[width=.3\textwidth]{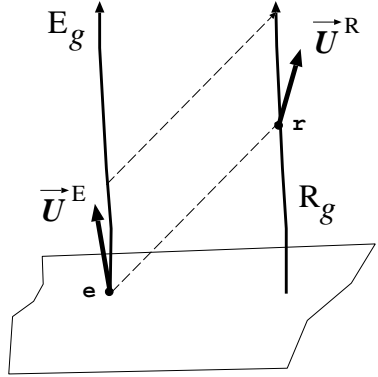}
\end{center}
\caption{Construction of the timelike geodesic worldlines E$_{\rm g}$
and R$_{\rm g}$, and of the Minkowski-like coordinates.  }
\label{fig:appendixfig}
\end{figure}
%%%%%%%%%%%%%%%%%%%%%%%%%%%%%%%%%%%%%%%%%%%%%%

We now turn to a proof, needed in Sec.~\ref{sec:gaugeinv},  that the
expression in Eq.~(\ref{DoppwithU}) is the Doppler shift. 
Since we have shown that the expression in 
Eq.~(\ref{DoppwithU}) is gauge invariant, it suffices to show that
it is equal to Doppler shift in any one gauge.

We construct a convenient gauge for this proof as follows. We let {\bf
  e} and {\bf r} be the emission and reception events for a pulsar
photon. We choose a timelike geodesic worldline E$_{\rm g}$ through
{\bf e} to be almost tangent (i.e.\,, tangent to order $v $), at {\bf
  e}, to the emitter worldline.  Next we choose a timelike geodesic
worldline R$_{\rm g}$ through {\bf r} in such a way that it is
parallel to the geodesic worldline through {\bf e} in the following
sense.  Near the emission event we construct a spacelike surface
orthogonal to the emitter 4-velocity, and we construct this surface to
have extrinsic curvature with a vanishing trace.  We generate (much in
the manner of Gaussian normal coordinates) a congruence of timelike
geodesic worldlines normal to this surface. We assume that the
spacetime curvature is small enough, and/or the emitter-receiver
distance is small enough that this congruence fills spacetime, with no
crossings, in the neighborhood of the reception event. We choose
R$_{\rm g}$ to be the curve in that congruence that goes through {\bf
  r}.

The next step in the construction is to generate null geodesics
(photon worldlines) from E$_{\rm g}$ to R$_{\rm g}$ and to define our
Minkowski coordinates $t,x,y,z $ by the following steps: (i)~the
surface spanned by the null geodesics is taken to be a surface of
constant $x$ and $y$; (ii)~$z$ is set to zero along E$_{\rm g}$, and
along R$_{\rm g}$ we set $z $ equal to another constant, the length of
the spatial geodesic between E$_{\rm g}$ and R$_{\rm g}$ on the
spatial hypersurface. (iii)~$z$ is taken to be an affine parameter
along the null geodesics; (iv)~$t$ is taken to be proper time along
E$_{\rm g}$; (v)~the coordinate $t$ is propagated through the $t,z$
surface by requiring $t-z$ to be constant along the null geodesics, we
note that a tangent to the null geodesics is given by
\begin{equation}
\left.{\partial_t}\right|_{z=\rm
const}+\left.{\partial_z}\right|_{t=\rm const}\,,
\end{equation}
in the coordinate system we have defined.

We now note that the covariant $t$ component of the photon 4-momentum 
satisfies the geodesic equation
\begin{equation}
\frac{dP_t}{d\lambda}=
P_\alpha \frac{dx^\beta}{d\lambda}\Gamma^\alpha_{\beta t}
=\textstyle{\frac{1}{2}}
P^\alpha \frac{dx^\beta}{d\lambda}h_{\alpha\beta,t}
\,
\end{equation}
and 
\begin{equation}
\Delta P_t=\int_{\rm E}^{\rm R} P^\alpha{dx^\beta} h_{\alpha\beta,t}
=
P^t\int_{\rm E}^{\rm R}\left(\textstyle{\frac{1}{2}}h_{tt,t}+h_{tz,t}+\textstyle{\frac{1}{2}}h_{zz,t}
\right)\,d\lambda
\end{equation}
where
\begin{equation}
\frac{d}{d\lambda}=
\left.\frac{\partial}{\partial t}\right|_{z=\rm
const}+\left.\frac{\partial }{\partial z}\right|_{t=\rm const}\,.
\end{equation}

From this expression we have that the difference between the photon
energies at {\bf r} and {\bf e} is
\begin{equation}\label{DeltaE}
[\mbox{Energy measured on R$_g$}-\mbox{Energy measured on E$_g$}]/P^t=-
\int_{\rm E}^{\rm R}\left(\textstyle{\frac{1}{2}}h_{tt,t}+h_{tz,t}+\textstyle{\frac{1}{2}}h_{zz,t}
\right)\,d\lambda\,. 
\end{equation}
We now generalize this to the case in which the photon is not confined
to the plane with $x$ and $y$ constant. Since the expression on the
right in Eq.~(\ref{DeltaE}) is already first order in the metric
perturbations, we need to consider only a more general photon
direction in the Minkowski background. If we denote by $\vec{n}$ the
unit vector pointing in the spatial direction in which the photon
moves, then the generalization of Eq.~(\ref{DeltaE}) is
\begin{equation}\label{DeltaE2}
[\mbox{Energy measured on R$_g$}-\mbox{Energy measured on E$_g$}]/P^t=-
\int_{\rm E}^{\rm R}
\left(\textstyle{\frac{1}{2}}h_{tt,t}+n^jh_{tj,t}+\textstyle{\frac{1}{2}}h_{zz,t}
\right)\,d\lambda\,. 
\end{equation}

To get the full expression for the Doppler shift we must consider the
fractional energy changes from the geodesic worldlines to the observer
worldlines. An observer with 4-velocity $U^\mu$ observes a photon with
4-momentum $P^\mu$ to have energy $-P_\mu U^\mu$.  By construction,
our geodesic worldlines have components $U^\mu=\{1,0,0,0\}$, Thus the
energy observed by the receiver is $P^t\left[U^t-n^jU_j\right]^R$,
where $P^t$ is the energy observed at the reception event by the
geodesic observer. With this and the similar expression for the emission 
event we get Eq.~(\ref{DoppwithU}).

\bibliographystyle{apj}
%\bibliography{riemann}

\clearpage

\end{document}